\documentclass{ethpaper}
\usepackage{graphicx}
\usepackage{latexsym,rotating,subfigure}

\newcommand{\expfor}[2]{$#1\!\times\! 10^{#2}$}

\begin{document}

\begin{titlepage}

   \ethnote{}
\title{First results on radiation damage in PbWO$_4$ crystals exposed
to a 20\,GeV/c proton beam}

\begin{Authlist}
M. Huhtinen
\Instfoot{cern}{CERN, CH-1211 Geneva, Switzerland}
P.~Lecomte, D.~Luckey and F.~Nessi-Tedaldi
\Instfoot{eth}{Swiss Federal Institute of Technology (ETH),
CH-8093 Z\"urich, Switzerland}
  \end{Authlist}
\maketitle
\abstract{ We have exposed seven full length production quality
crystals of the electromagnetic calorimeter (ECAL) of the CMS detector
to a 20\,GeV/c proton beam at the CERN PS accelerator. The exposure
was done at fluxes of $10^{12}$\,p/cm$^2$/h and
$10^{13}$\,p/cm$^2$/h and integral fluences of
$10^{12}$\,p/cm$^2$ and $10^{13}$\,p/cm$^2$ were reached at both
rates. The light transmission of the crystals was measured after
irradiation and suitable cooling time for induced radioactivity to
decrease to a safe level. First results of these measurements are
shown.  The possible damage mechanisms are discussed and simulations
based on one possible model are presented. The implications for
long-term operation of CMS are discussed and it is shown that in the
whole barrel and at least most of the ECAL endcap hadron damage alone
-- even if cumulative -- should not cause the crystals to fail the CMS
specification of $\mu_{\rm IND}<1.5$\,m$^{-1}$ during the first
10\,years of LHC operation.}
\vspace{7cm}
\conference{Presented at the {\it8th ICATPP Conference\\ on Astroparticle, Particle,
Space Physics, Detectors and Medical Physics Applications}\\ Como, Italy,
6 to 10 October 2003}

\end{titlepage}

\setcounter{page}{2}
\section{Introduction}
\label{s-introduction}
Lead Tungstate (PbWO\,$_4$) crystals are being used by several
high-energy physics experiments because they provide a compact
homogeneous calorimeter with fast scintillation. In view of the harsh
radiation conditions expected in the CMS experiment, the radiation
hardness of PbWO\,$_4$ has been extensively studied with
$\gamma$-irradiations.  A systematic irradiation study with
high-energy hadrons has so far not been extended up to the full
fluences expected at the LHC.

The fundamental difference between $\gamma$-irradiation and energetic
hadrons is that the latter produce inelastic nuclear interactions
(stars) in the crystal. These interactions break up the target nucleus
and thus create impurities and distortions in the crystal lattice.
The slow nuclear fragments produced in each hadronic interaction can
have a range of up to 10\,$\mu$m. Along their path they displace a
large number of lattice atoms, but also ionize much more densely than
a minimum ionizing particle. Neither the displacements, nor the dense
ionization, although in a very small volume, can be produced in
$\gamma$-irradiations.  Since the threshold for star production is
$\sim$20\,MeV, reactor neutron irradiations do not cover this regime
either.
\section{The crystals, irradiations and measurements}
\label{s-crystals}
We have studied CMS production quality crystals from Bogoroditsk.  The
crystals have a nearly parallelepipedic shape with dimensions of
$2.6\times 2.6\times 23$\,cm$^3$.  Except for slightly non-compliant
mechanical dimensions all crystals satisfied the technical
specifications for their use in the CMS ECAL. All crystals were
pre-tested for radiation hardness at the Geneva Hospital
$\gamma$-irradiation facility for 2\,h at 250\,Gy/h and showed a
radiation induced absorption coefficient $\mu_{\rm IND}<1.5$\,m$^{-1}$
at 440\,nm as required by the CMS technical
specifications~\cite{r-ECALTP}. The crystals were subsequently
annealed.

An irradiation test of PbWO\,$_4$ with hadrons is complicated mainly
by the fact that the crystals become highly radioactive. Therefore our
aim was to make the simplest possible test by irradiating bare
crystals and measuring the longitudinal light transmission (LT) after
irradiation.
\begin{figure}
\begin{minipage}[t]{0.46\textwidth}
\setlength{\unitlength}{1cm}
\vspace{-40mm}
{\footnotesize
Table\,1: Fluences determined from $^{22}$Na activity in the Al foils and
total fluences used for our analysis.\vspace*{1pt}}
\begin{picture}(10.0,4.0)
\put(0.4,1.7){
{\footnotesize
\begin{tabular}{|c|c|c|} \hline
ID      & $\Phi$($^{22}$Na)   & $\Phi$-total        \\ \hline
a       & \expfor{9.64}{11}   & \expfor{9.64}{11}   \\
F       & \expfor{1.54}{12}   & \expfor{1.54}{12}   \\
a'      & \expfor{1.37}{13}   & \expfor{1.47}{13}   \\
b       & \expfor{5.21}{11}   & \expfor{5.21}{11}   \\
c       & \expfor{8.36}{11}   & \expfor{8.36}{11}   \\
d       & \expfor{1.35}{13}   & \expfor{1.35}{13}   \\
E       & \expfor{8.34}{12}   & \expfor{8.34}{12}   \\
F'      & \expfor{9.10}{12}   & \expfor{1.67}{13}   \\
G       & \expfor{1.93}{12}   & \expfor{1.93}{12}   \\ \hline
\end{tabular}}}
\end{picture}
\end{minipage}
\hspace{0.05\textwidth}
\begin{minipage}[t]{0.46\textwidth}
\setlength{\unitlength}{1cm}
 \begin{picture}(10,4.0)
   \put(-0.2,-0.3){\mbox{\includegraphics[height=50mm]{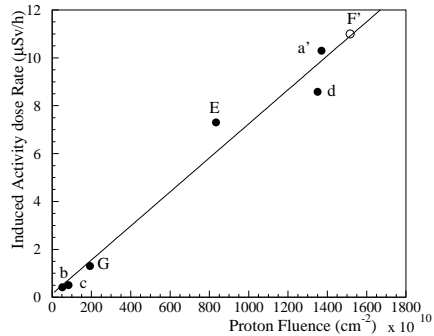}}}
 \end{picture} 
\caption{
Correlation between fluence from $^{22}$Na activity and induced activity of the
crystals. For F', see the explanations in the text.} 
\label{actflu}
\end{minipage}
\end{figure}
\addtocounter{table}{1} 
The irradiations were performed at the IRRAD1
facility in the T7 beam-line of the CERN PS accelerator.  The proton
momentum was 20\,GeV/c and the exposure was uniform to within a factor
of two over the crystal front face at fluxes of
$\sim10^{12}$\,p/cm$^2$/h and $\sim10^{13}$\,p/cm$^2$/h. The beam was
parallel to the long axis of the crystal. The fluence was determined
by $^{22}$Na activity in aluminum foils in front of the crystals and
correlated with the induced radioactivity in the crystals themselves
measured at $4.5$ cm distance, as shown in Fig.\,\ref{actflu}. The
$^{22}$Na activity for the irradiation of crystal F' is inconsistent for
some unknown reason. For this crystal we
deduce the fluence by using the induced radioactivity and the
correlation determined from the other irradiations. Table\,1 shows the fluence
from $^{22}$Na and the total fluence used in our analysis.  Crystals
which were irradiated at a rate of $10^{12}$\,p/cm$^2$/h are
indicated by small letters, those exposed to
$10^{13}$\,p/cm$^2$/h by capital letters. Two of the crystals,
labeled a(a') and F(F'), were irradiated twice, so for a' and F' the
last column of table\,1 shows the added fluence of two irradiations.

We have measured LT for each crystal with a Perkin Elmer Lambda 900
spectrophotometer using depolarized light of 300--800 nm in 1 nm
steps. The light beam was about 7\,mm wide and 10\,mm high. A
check of reproducibility showed that the measured LT has an accuracy
of $\pm$\,1\%~\cite{r-LTPREC}.
\section{Results}
\begin{figure}[htb]
\begin{center}\footnotesize
  \begin{tabular}[t]{cc}
\subfigure[Longitudinal Transmissions after one low and one high fluence
irradiation for crystals ``a'' ($10^{12}$ p/cm$^2/h$) and ``F'' 
($10^{13}$ p/cm$^2/h$). \label{f-AandFLT}]
{\includegraphics[width=0.45\textwidth]{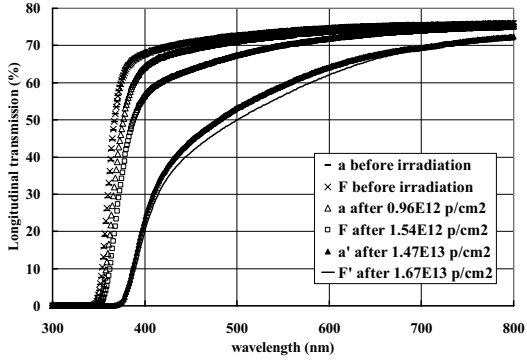}}&
\subfigure[Induced absorption coefficients at 440\,nm for all
crystals, measured 20 and 50 days after irradiation, as a function of total
fluence.\label{f-muvsfluence}]
{\includegraphics[width=0.45\textwidth]{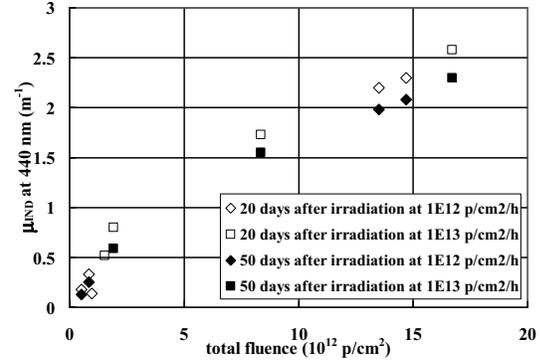}}
\end{tabular}
\end{center}
\caption{Longitudinal Transmission changes dependence on hadron fluence.}
\label{f-results}
\end{figure}
Fig.~\ref{f-AandFLT} shows the full LT curves over the whole measured
range of wavelengths, for crystals a(a') and F(F').  Crystal ``a'' was
irradiated at a 10 times smaller rate than crystal ``F''. As can be
seen, the LT for the lower fluence irradiation is different between
``a'' and ``F''. This can be due to the fact that $\mu_{\rm
IND}(440\,{\rm nm})$ might be dominated by the damage induced by the
total ionization, i.e. conditions similar to
$\gamma$-irradiations. Such a damage is known to saturate at a level
which depends on the dose rate. At the higher fluences, however, the
LT curves are quite similar, which might indicate that between
integral fluences of $10^{12}$\,p/cm$^2$ and $10^{13}$\,p/cm$^2$
specific hadronic damage starts to dominate. An important observation
is also that after hadron exposure the LT band-edge has shifted to longer
wavelengths.

Fig.\,\ref{f-muvsfluence} shows, as a function of fluence,
$\mu_{\rm IND}(440\,{\rm nm})$ measured 20 and 50 days after the end
of each irradiation. At low fluences our data do not allow to make a
distinction between a pure dose rate and a possibly cumulative
effect. At high fluences the dependence on rate is very small, while
our data indicate a clear -- possibly linear -- dependence on integral
fluence.
\begin{figure}  
\begin{center}                   
\includegraphics[height=5.5cm]{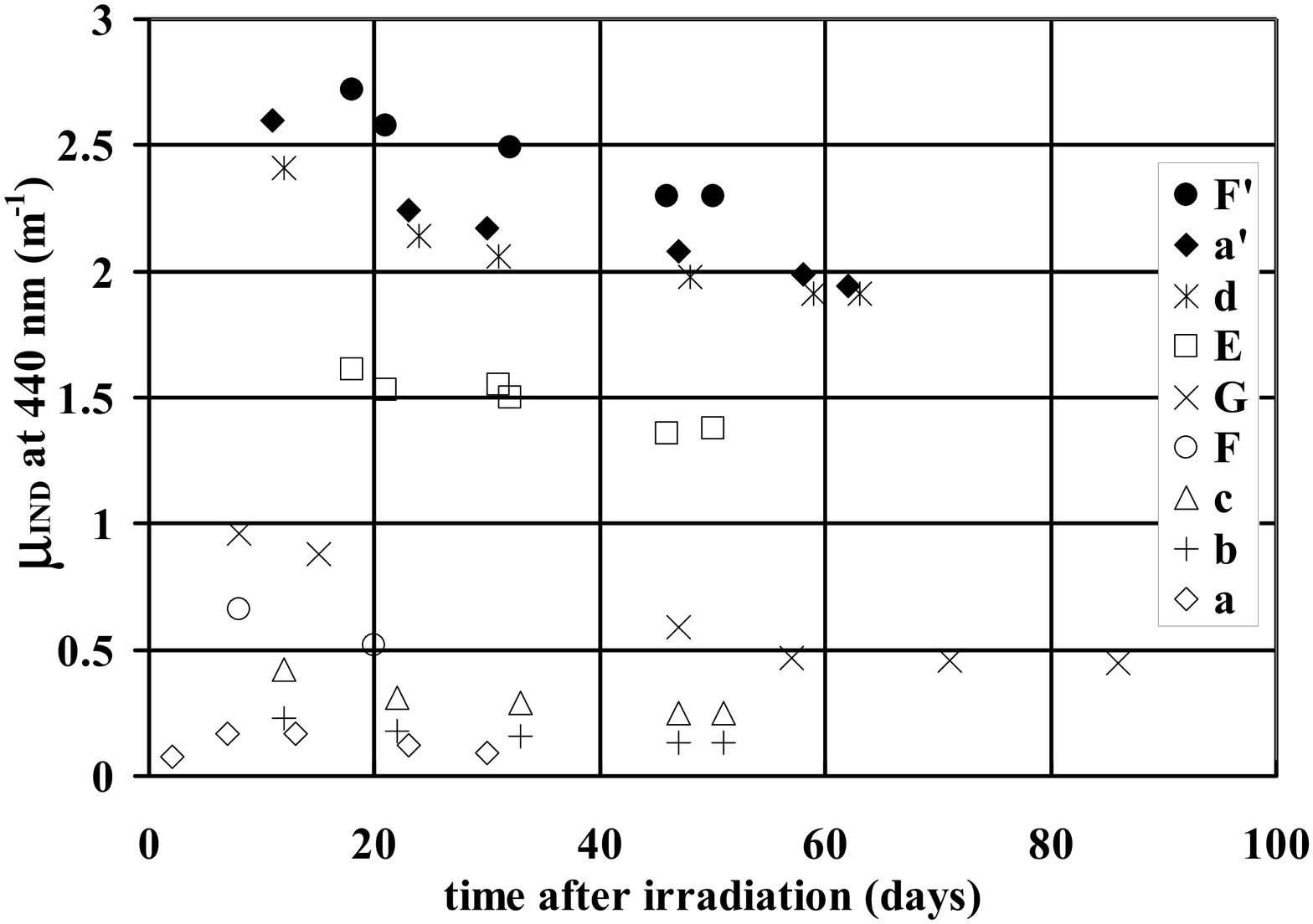}
\caption{Crystal recovery. Because crystals irradiated to
$\sim 10^{13}$\,p/cm$^2$ 
could be measured only 10 days after irradiation, short time constants are not
visible.}
\label{muvsdays}
\end{center}
\end{figure}
With the fluences reached and for the rates used, we observe no sign
of saturation.
Causes of the effects observed could be:\\
- High-energy hadrons
producing a specific cumulative damage, with essentially no recovery,
combined with a damage which anneals at room temperature with a time
constant of at least several months.\\
- The $\gamma$ dose rate caused
by the charge of the protons~\cite{tvirdee}.

Room temperature recovery, recorded up to now, is shown in
Fig.\,\ref{muvsdays}. The data do not yet allow to extrapolate to
LHC-like time-scales.
\section{Monte-Carlo simulations}
Simulations were performed to understand the hadron fluences, star
density and dose as a function of depth in the crystal. They show that
after an initial increase all distributions are fairly flat with a
slow decrease after a maximum at $\sim$7.5\,cm.\\

\begin{figure}[htb]
\begin{center}
\setlength{\unitlength}{1cm}
\begin{picture}(10,6)
\put(0.0,0.0){
\includegraphics*[height=7cm]{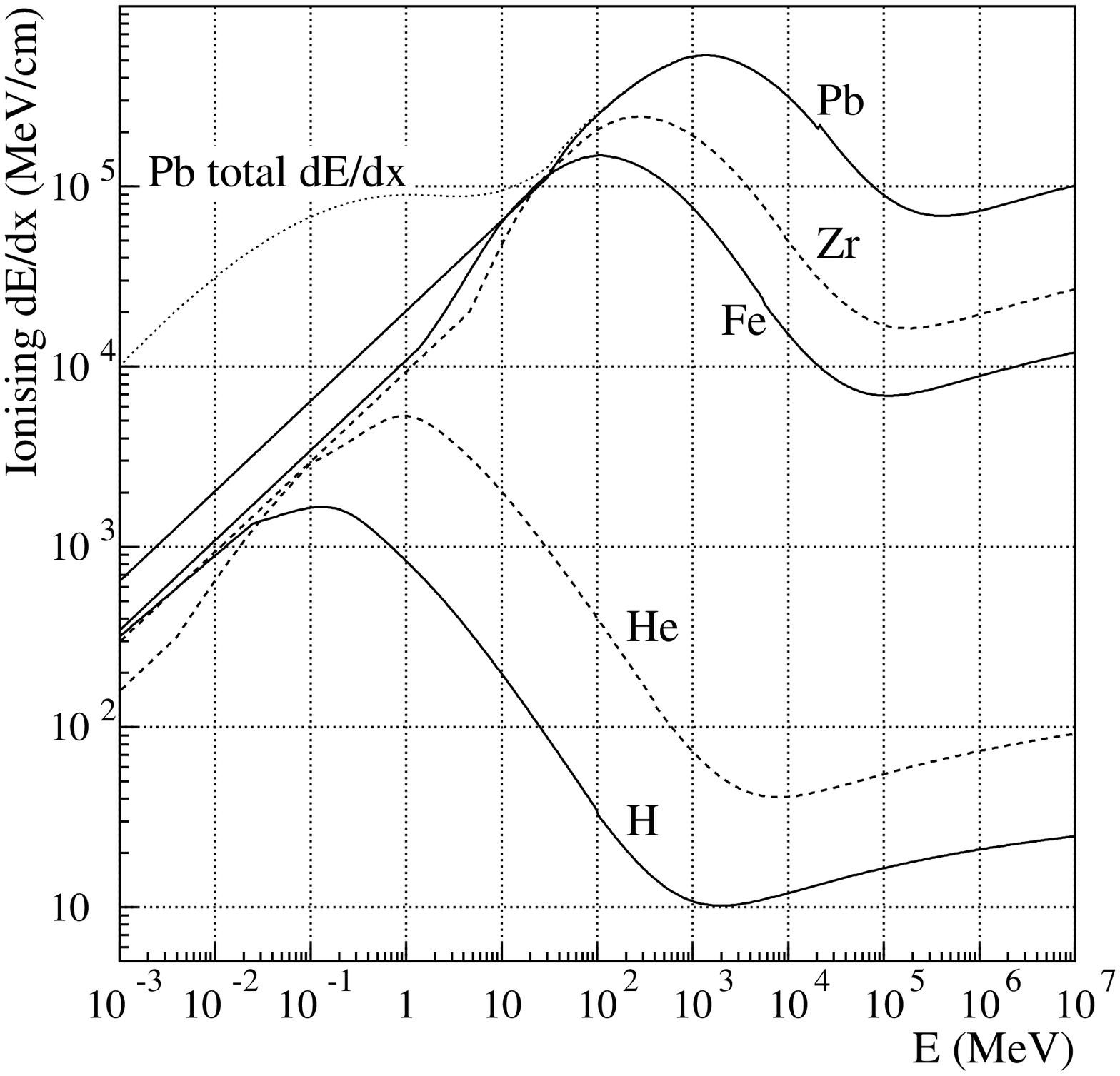}}
\put(6.8,0.0){
\includegraphics*[height=7cm]{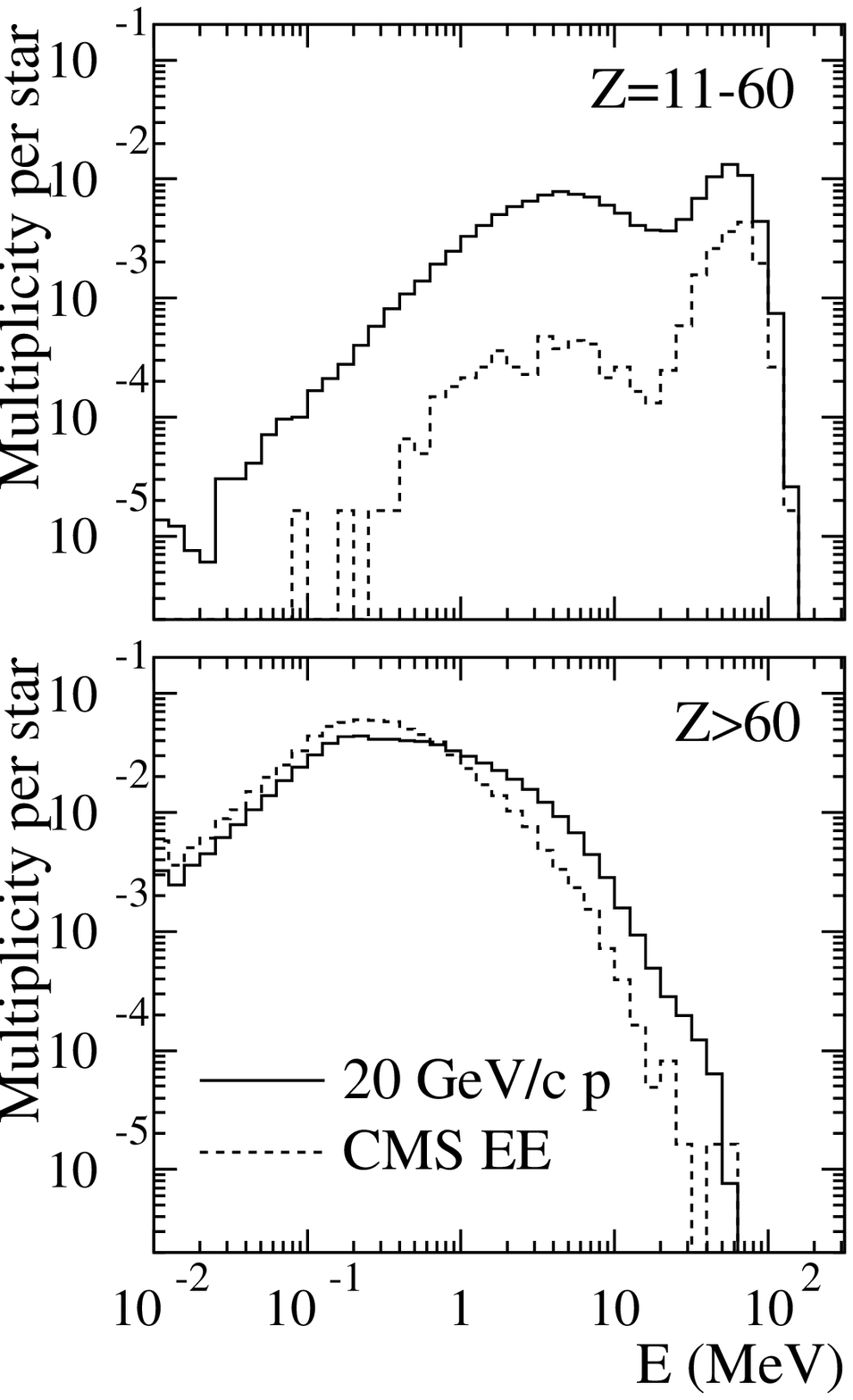}}
\end{picture} 
\caption{Ionizing dE/dx (and total for Pb) of ions in PbWO$_4$ and simulated
fragment spectra from hadronic interactions.}
\label{dedx}
\end{center}
\end{figure}
\begin{figure}
\begin{center}
\setlength{\unitlength}{1cm}
\begin{picture}(10,5)
\put(0.8,-0.3){
\includegraphics*[height=60mm]{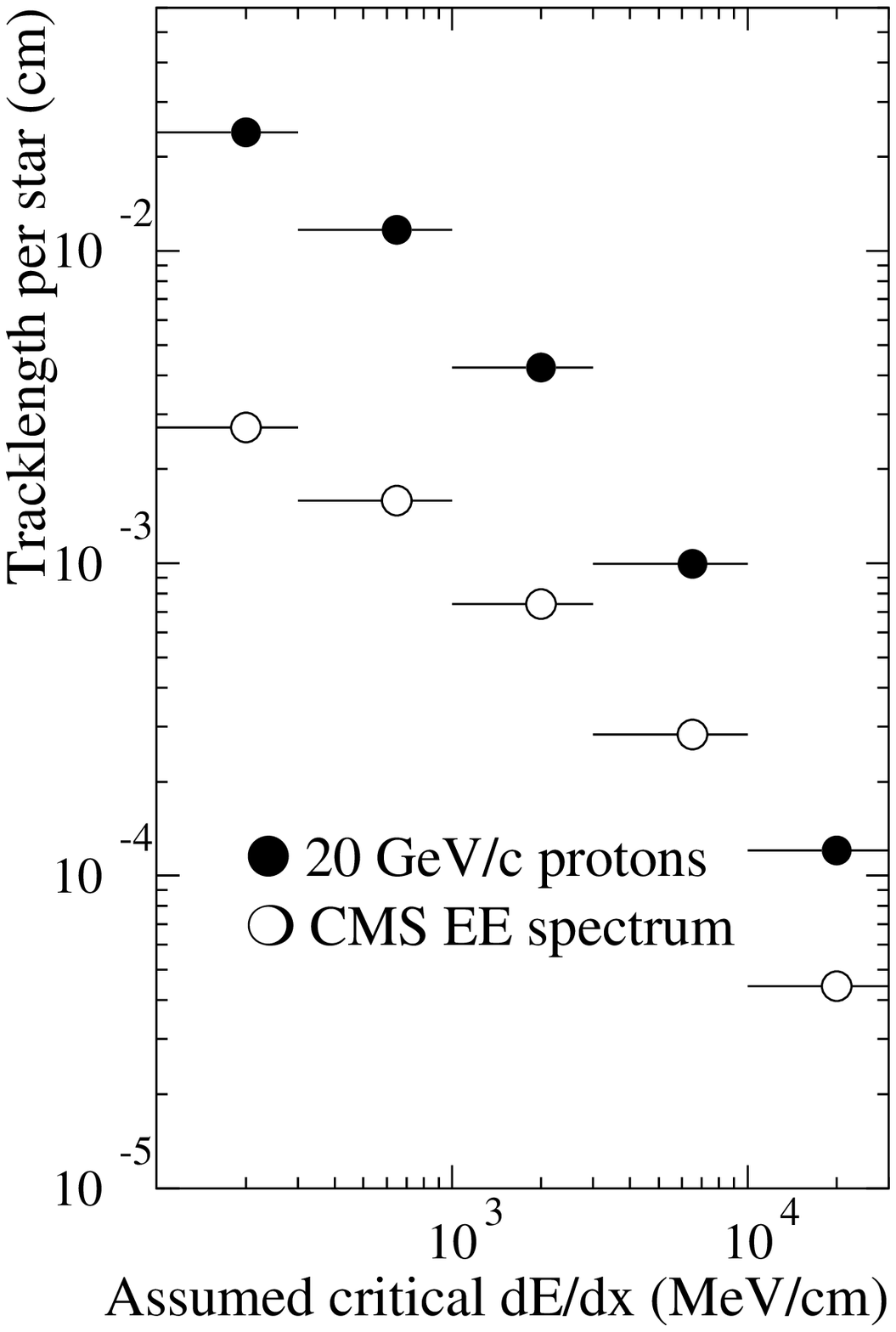}}
\put(5.0,-0.3){
\includegraphics*[height=60mm]{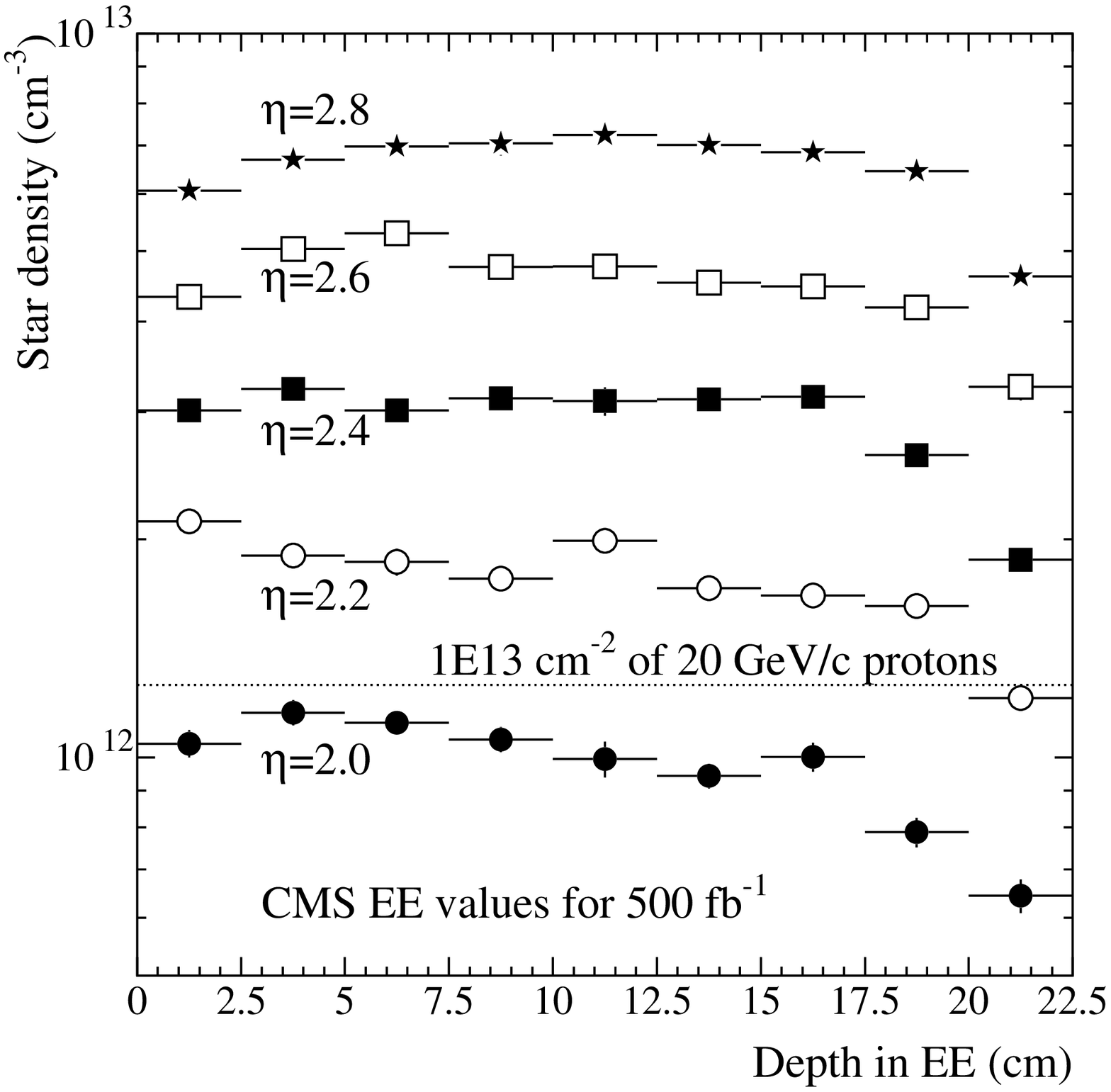}}
\end{picture} 
\caption{Track length above a given (dE/dx)$_{\rm crit}$ per star (inelastic hadronic interaction) 
and star densities for 10 years in the ECAL endcap and for our irradiation test.}
\label{dedxpstar}
\end{center}
\end{figure}

The average values for one proton per cm$^{2}$ are 1.63\,cm$^{-2}$ for
the charged hadron fluence, 0.126\,cm$^{-3}$ for the star density and
1.37\,nGy for ionizing dose.
A detailed simulation of the full atomic cascade initiated by the
fragments shows that for an incident proton fluence of
$10^{13}$\,p/cm$^2$ the concentration of displacements and
interstitials might reach $10^{-8}$ which is far below the
pre-irradiation imperfections.  Thus if specific hadronic damage is
produced, it is more likely to be due to the dense ionization of the ion
tracks. Fig.\,\ref{dedx} shows that for the fragment spectra produced
by inelastic interactions, the energy loss can be more than 4 orders
of magnitude above the dE/dx of a minimum ionizing particle. Making
the {\em hypothesis} that tracks exceeding a value (dE/dx)\,$_{\rm
crit}$ are responsible for hadronic damage, we can compare the
20\,GeV/c irradiation with the expected CMS ECAL endcap (EE)
conditions. Fig.\,\ref{dedxpstar} shows that, depending on
(dE/dx)$_{\rm crit}$, the simulated track length per star produced by
the 20\,GeV/c beam is a factor between 2.7 and 8.7 higher than in the
EE. The star density for $10^{13}$\,cm$^{-2}$ protons corresponds to
more than $\eta$=2 in the EE. With the possible factor of 2.7--8.7,
our irradiations cover the high-precision area of the EE
for 10\,years operation and possibly even the whole EE up to
$\eta$=2.9.
\section{Conclusions}
Our LT measurements of PbWO$_4$ crystals after exposure to 20\,GeV/c
protons are consistent with a hypothesis of cumulative hadronic
damage, although the small number of crystals and limited fluence
range does not allow a definitive statement yet.  Room temperature
recovery is very slow and possibly partial.
Even under the most pessimistic assumption of permanent cumulative
damage, our results verify that the CMS barrel ECAL will meet its
design specifications over 10 years. Simulations, used to compare the
conditions in our test beam and the EE under the hypothesis of
hadronic damage, indicate that in most of the EE $\mu_{\rm IND}$ due
to hadronic damage alone is likely to stay below 1.5\,m$^{-1}$ for
10\,years of LHC operation.  To gain a better understanding we will
extend our irradiations to higher fluences (\expfor{5}{13}\,p/cm$^2$)
and to possibly lower hadron energies, and we will follow the damage
recovery. Finally, comparative $\gamma$-irradiations at dose rates
corresponding to those of the proton beam are expected to provide the
most stringent test of specific hadronic damage.
\section*{Acknowledgments}
These irradiations were made possible only by the efforts of
R. Steerenberg, who provided us with the required beam conditions. The
help of M. Glaser and F. Ravotti in operating the irradiation and
dosimetry facilities is gratefully acknowledged. The support of
T.Virdee and F.Pauss has been essential. We thank R.M.Brown for
stimulating discussions.
\end{document}